\documentclass[aps,pra,showpacs,superscriptaddress,12pt]{revtex4}%
\usepackage{amsmath,amssymb,graphicx,epsfig}
\usepackage{bm}
\usepackage{graphicx}

\begin{document}

\title{A critical analysis of Popper's experiment}

\author{GianCarlo \surname{Ghirardi}}\email{ghirardi@ts.infn.it}%
\affiliation{Department of Theoretical Physics, University of
  Trieste, Italy}%
\affiliation{Istituto Nazionale di Fisica Nucleare, Sezione di Trieste,
  Italy}%
\affiliation{International Centre for Theoretical Physics ``Abdus Salam,''
  Trieste, Italy}%

\author{Luca \surname{Marinatto}}\email{marinatto@ts.infn.it}%
\affiliation{Department of Theoretical Physics, University of
  Trieste, Italy}%
\affiliation{Istituto Nazionale di Fisica Nucleare, Sezione di Trieste,
  Italy}%

\author{Francesco \surname{de Stefano}}\email{francesco.destefano@alice.it}%
\affiliation{Liceo Scientifico ``G. Marinelli", Udine, Italy}%

\date{\today}

\begin{abstract}
An experiment which could decide against the Copenhagen interpretation of quantum mechanics has been proposed by
K. Popper and, subsequently, it has been criticized by  M.J. Collett and R. Loudon. Here we show that both the
above mentioned arguments are not correct because they are based on a misuse of  basic quantum rules.
\end{abstract}

\maketitle


\section{Introduction}

An experiment involving a couple of entangled particles emitted in opposite directions from a common source has
been proposed by K.Popper~\cite{popper} as a crucial test for the Copenhagen interpretation of quantum
mechanics. We begin by presenting the problem  in the  terms used by Popper and by Collett and
Loudon~\cite{collett}, and only subsequently we will be more precise about it.
\begin{figure}[h]
\begin{center}
\includegraphics[scale=0.5]{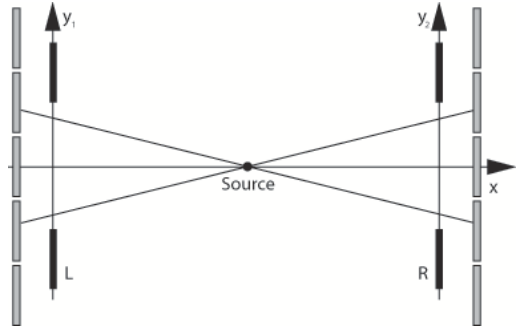} \caption{The picture shows the initial arrangement of
Popper's proposed experiment. The black vertical lines determine the width of the slits, the sequence of
vertical gray boxes represents the array of counters behind the slit.} \label{fig1}
\end{center}
\end{figure}
The two particles are assumed to propagate (see Fig.~\ref{fig1}) along the $x$-axis towards two arrays of
detectors wired to operate in coincidence and placed at  left (L) and right (R) of the emitting source,  at
equal distances from it.

Two slits orthogonal to the flight direction are placed, along the $y$-axis,  before an array of detectors and,
initially, their width $\Delta$ is wide enough so that only the counters which are placed directly behind them
get activated. In fact, if the wave function of the particles is suitably prepared, only the detectors placed at
small angles with respect to the flight direction $x$ will register the arrival of a particle with a
non-negligible probability. Subsequently,  the slit at R is narrowed so as to produce an uncertainty-principle
scatter which appreciably increases the set of counters behind the slit which may be activated with a
non-negligible probability. Popper's argument~\cite{popper} goes then as follows: if the Copenhagen
interpretation of quantum mechanics is correct, any increase in the precision of the knowledge of the
$y$-position of the particle at R should correspond to an analogous increase of the knowledge concerning the
$y$-coordinate of the particle at  L, as a consequence of the assumed strict correlations implied by the
entanglement. Hence, also the scatter of the particle at L should instantaneously increase, even if the width of
the slit at this side has not been modified. This prediction is testable, since new counters would be activated
with an appreciable probability,  giving rise to a detectable form of superluminal influence at odds with the
postulates of special relativity.

\begin{figure}[h]
\begin{center}
\includegraphics[scale=0.5]{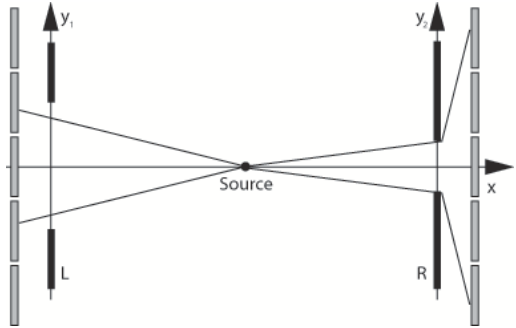} \caption{The analogous of Fig. 1 for the case in which the
slit at R has been remarkably narrowed. After the slit the scatter in position increases remarkably.}
\label{fig2}
\end{center}
\end{figure}

To save the principle of causality, Popper concludes, no wider scatter at the left hand side L will occur by
narrowing the slit at R (see Fig.~\ref{fig2}).
As a consequence, the Copenhagen interpretation of quantum mechanics which, it has to be stressed, in Popper's
view predicts a wider scatter at L induced by the ``mere knowledge"~\cite{popper} of the position of the
particle at $L$ which has been caused by the narrowing of the slit at R, would be falsified. In brief, either
the Copenhagen interpretation or special relativity must be abandoned, and Popper  definitively  favors  the
first option.

Subsequently, a criticism of Popper's proposal by Collett and Loudon appeared~\cite{collett} (along with an
exchange of views~\cite{replies}), claiming  that the experiment cannot represent  a crucial test for the
Copenhagen interpretation of quantum mechanics. The reason for this, in the authors' opinion, is that the
non-negligible indeterminacy about the state of the source would make undetectable the effects which the
Copenhagen interpretation predicts and Collett and Loudon fully agree with Popper about such an interpretation.
In fact, in Ref.~\cite{collett} these authors pretend to prove that the limitations imposed by the uncertainty
principle on the accuracy with which the state of the source can be specified, unavoidably remove the increase
in scatter induced by the (indirect) more precise knowledge of the position of the particle at L. Even more, the
authors surprisingly claim (as it is put into clear evidence also by Eq.~(9) and by Fig.~2 of their
paper~\cite{collett}) that both the width of the slit at R and the amplitude of the scatter at L are expected to
decrease together. To conclude, Collett and Loudon  agree with Popper's statement about the alleged implications
of the Copenhagen interpretation of the proposed test, and address their criticism solely to the unphysical
assumption that the source is stationary at a fixed and known position.

However, as we are going to prove here, the situation is radically different from the one described in the above
mentioned papers: the argument developed by Popper~\cite{popper} is invalid not because of some unjustified
assumption about the source but, rather, because the (universally accepted) rules of quantum mechanics, even in
the Copenhagen interpretation, do not entail what Popper claims them to entail.

In this paper we will resort to two (equivalent) arguments showing clearly the fallacy of Popper's idea. The
first one descends from a well-known theorem on the impossibility of any form of superluminal signaling using
the quantum correlations~\cite{ghirardi}. In fact, if one could induce an instantaneous larger scatter of a
particle at L by simply narrowing a slit at place R (as described in Popper's test) taking advantage of the
peculiar spatial correlations of entangled wave functions, a superluminal signaling protocol between two
arbitrarily distant parties would have been achieved. However, as we will see, quantum formalism forbids in
complete generality such a possibility, guaranteeing  the peaceful coexistence~\cite{shimony} with the causality
principle dictated by special relativity. The second argument, which we include for pedagogical reasons, shows
by explicit calculations that, given a (particular and simple) entangled quantum state which correctly describes
the kind of spatial correlations envisaged by Popper, it is impossible to induce a larger scatter at position L
by narrowing the slit at R. It is almost amusing to observe that our precise analysis makes absolutely clear
that a correct use of the quantum rules implies exactly what Popper was waiting for (i.e., no larger scatter at
L).

Both our arguments, the first one being completely general while the second being based on direct calculations
performed on a particular, though representative, quantum entangled state, will definitely show the reasons why
Popper's argument~\cite{popper}, and the ensuing discussions~\cite{collett}, cannot be considered crucial
neither for the Copenhagen interpretation of quantum mechanics nor for the causality principle of special
relativity. Moreover we hope that our analysis will contribute to stop the investigations and even the proposals
of actually performing the experiment which, from time to time, reappear in the literature~\cite{experiments}.


\section{No signaling condition}

As already stated, it is a remarkable feature of quantum mechanics its peaceful coexistence with special
relativity. In fact, within a quantum mechanical framework, it is impossible to devise a superluminal signaling
protocol between two distant parties by resorting to the nonlocal correlations exhibited by entangled states.
This is due to the fact that (i) arbitrary unitary operations applied locally on a particle cannot alter the
probability distributions of measurement processes performed onto the other particle, (ii) the genuine
stochasticity of the reduction process prevents one to control the measurement outcomes, independently both from
the kind of measurements one actually performs (it holds true both for projection valued~\cite{ghirardi} and
positive operator valued measurements~\cite{ghirardi2}) and from the choice of the specific reduction mechanism
which is assumed to govern the measurement process (wave packet reduction, spontaneous localization or any other
conceivable reduction process). This result is commonly referred to as {\em no superluminal signaling condition}
in quantum mechanics. For the sake of brevity in this section we will present a simplified version of the no
signaling theorem by limiting our consideration only to the case of local unitary operations and of measurements
described by projection operators within standard quantum mechanics, and address the reader to
Ref.~\cite{ghirardi2} for the most general scenario.

To this end, consider a composed bipartite quantum system described by an arbitrary (normalized) statistical
operator $\rho_{12}$ acting on the finite dimensional Hilbert space ${\cal{H}}_{1}\otimes {\cal{H}}_{2}$. Now,
suppose that an observer performs locally an arbitrary unitary operation, represented by the operator $U_{2}$,
onto particle $2$. Accordingly, the initial state of the system $\rho_{12}$ becomes
${\tilde{\rho}}_{12}=I_{1}\otimes U_{2} \rho_{12} I_{1}\otimes U^{\dagger}_{2}$. This local action does not
modify the reduced statistical operator associated to particle $1$ since
\begin{equation}\label{eq0.01}
{\textrm{Tr}}^{(2)}[{\tilde{\rho}}_{12}] =  {\textrm{Tr}}^{(2)}[I_{1}\otimes U_{2} \rho_{12} I_{1}\otimes
U^{\dagger}_{2} ]  = {\textrm{Tr}}^{(2)}[\rho_{12} I_{1}\otimes U^{\dagger}_{2}U_{2}
]={\textrm{Tr}}^{(2)}[\rho_{12}]
\end{equation}
where the cyclic property of the (partial) trace operation has been used. Thus, no unitary operation applied
locally to one of the two particles can alter the probability distributions of measurements performed onto the
other particle. Now, it remains to be proven that also a local nonselective projective measurement cannot be
effective for sending a superluminal message. To this end, let us consider two Hermitian operators
$A=\sum_{i}a_{i}P_{a_{i}}$ and $B=\sum_{j}b_{j}Q_{b_{j}}$, referring to the first and to the second component
subsystems respectively. For simplicity, we suppose that the spectrum $\{ a_{i}\}$ and $\{b_{j}\}$ of both
obervables is nondegenerate. In the previous expressions of the observables we have denoted as $\{P_{a_{i}}\}$
and $\{Q_{b_{j}}\}$ the associated family of (one-dimensional) orthogonal projection operators. Now, suppose
that initially an observer performs a (nonselective) measurement of the observable $B$ on his particle $2$
obtaining the outcome $b_{j}$: this causes the state $\rho_{12}$ of the system to collapse onto the (trace one)
state $I\otimes Q_{b_{j}}\rho_{12}I\otimes Q_{b_{j}}/{\textrm{Tr}}[I\otimes Q_{b_{j}}\rho_{12}]$ with
probability $Pr(B=b_{j})={\textrm{Tr}}[I\otimes Q_{b_{j}}\rho_{12}]$. Subsequently, a measurement of the
observable $A$ is performed onto the (arbitrarily distant) particle $1$ and we calculate the (nonselective)
probability distribution $Pr(A=a_{r}| B)$, ignoring the outcome which has been obtained in the previous
measurement process of $B$, as follows
\begin{eqnarray}
\label{eq0.1} Pr(A=a_{r}|B)& = & \sum_{j} Pr(A=a_{r}|B=b_{j}) Pr(B=b_{j}) \\
\label{eq0.2}  &= & \sum_{j} {\textrm{Tr}}\Big[\frac{P_{a_{r}}\otimes Q_{b_{j}}\rho_{12}P_{a_{r}}\otimes
Q_{b_{j}}}{{\textrm{Tr}}[I \otimes Q_{b_{j}}\rho_{12}]}
\Big]{\textrm{Tr}}[I\otimes Q_{b_{j}}\rho_{12}]\\
\label{eq0.3} & = & \sum_{j} {\textrm{Tr}}[P_{a_{r}}\otimes Q_{b_{j}}\rho_{12}] = {\textrm{Tr}}[P_{a_{r}}\otimes
I\rho_{12}]
\end{eqnarray}
where we have used the linearity and the cyclic property of the trace, the fact that operators belonging to
different spaces commute and that the family of projectors $\left\{Q_{b_{j}}\right\}$ constitutes a resolution
of the identity. Now, if we calculate the outcome probability distribution $Pr(A=a_{r})$ of the observable $A$
when no previous measurement of $B$ is performed, we easily notice that it coincides with the expression in
Eq.~(\ref{eq0.3}), that is, $Pr(A=a_{r}|B)=Pr(A=a_{r})$. This proves that no projective measurement procedure
performed on particle $2$ may affect the probability distributions of measurements performed subsequently on a
distant particle $1$, and this holds true irrespective of the state $\rho_{12}$, which might be entangled or
not.

To summarize, the previous arguments (and the more general result of~\cite{ghirardi2}) imply that local
operations performed on particle $2$ (such as unitary operations or projective measurements) cannot be used to
transmit any information whatsoever in the region where particle $1$ is located, in spite of the fact that
quantum correlations are genuinely nonlocal and the reduction mechanism is assumed to change instantaneously at
a distance the state vector of particle 2.

It should now be absolutely clear that the kind of experiment proposed by Popper~\cite{popper}, where an
observer placed at R might affect, at his will, the probability distribution of the $y_1$-component of the
momentum of the particle at L by simply narrowing a slit at R (i.e., by performing a $y_2$-position measurement
of the particle at R), actually contradicts the quantum mechanical rules. In fact, the unfolding of the process
hypothesized by Popper violates the theorem we have proved above~\footnote{A similar argument which criticizes
Popper's test has been developed in the paper by E. Gerjouy, A. Sessler, Am. J. Phys. {\bf 74}, 643 (2006).}.

Before concluding this section, a further remark is at order. We recall that, both in the original
proposal~\cite{popper} and in its subsequent analysis~\cite{collett}, the authors assume that the counters used
for the positions measurements of the particles are wired to operate in coincidence; in other words, only those
events in which two counters fire simultaneously at R and at L are considered for the test, while events in
which one of the particle gets absorbed by the slit (while the other does not) are discarded. Obviously, in this
specific situation, no observer at L could ever (not even hypothetically) think to get evidence of a
(superluminal) signal as a consequence of a (simultaneous) action performed at R (e.g., the narrowing of the
slit), simply because the two observers should know which events have to be recorded and which ones discarded.
Therefore, as it has been presented in Ref.~\cite{popper} and~\cite{collett}, Popper's test is, from its
starting hypothesis, completely meaningless for what concerns the possibility of exhibiting a violation of
special relativity.


\section{A pedagogical example}

In order to stress once again the fallacy of Popper's proposal without relying on general no-go theorems of the
kind exhibited in the previous section but by making explicit calculations, we shall resort, for pedagogical
reasons, to a mathematically unsophisticated modelization of Popper's experimental set up, which retains and
enlightens the distinctive features of the original proposal. These features comprise (i) the existence of
strict spatial correlations between the two particles, (ii) the fact that the wave function is initially chosen
so that only a limited fraction of counters are activated with a non-negligible probability when the slits are
wide-opened, and (iii) the occurrence of a detectable increase in scatter of a particle whenever it is passed
through a very narrow slit,  implying that it will activate counters at larger angles.

It is worth stressing that all these requests are indeed crucial in order that the test could be significative
and meaningful but, unfortunately, Popper's paper~\cite{popper} lacks of the needed clarity concerning precisely
these points. In fact, he does not make precise the mathematical form of the initial wave function of the
particle, limiting himself to speak of  an Einstein-Podoloski-Rosen-like state~\cite{epr} which should imply
perfect correlations between the positions $y_{1}$ and $y_{2}$ of the two particles. It is important to stress
that if this would really be the case, the wave function, at  time t=0 at which they reach the region where the
slits are placed, would  be  $\delta(y_{1}+y_{2})$, which is a non-normalizable state which does not fulfill
either of the requests (ii) and (iii). In fact it is easy to prove that the evolved wave function of such a
function, at any subsequent time, gives rise to a constant spatial probability distribution for both coordinates
$y_1$ and $y_2$. This amounts to say that perfect correlations in position necessarily imply that all counters
at both sides are activated with the same probability independently of the widths of the slits and, as a
consequence, that no further scatter will be ever detected when one narrows a slit. In accordance with these
remarks, one must release the condition of perfect correlations in order that the spread of the wave function be
not infinite and  Popper's test be meaningful.

\begin{figure}[h]
\begin{center}
\includegraphics[scale=0.5]{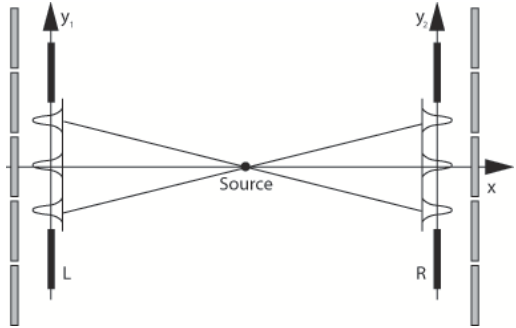} \caption{The explicit example of a correct Popper-like
experiment we are discussing. We have represented explicitly the three correlated Gaussians appearing in the
wave function. One has to keep in mind that we have assumed that the width $\sigma$ of these Gaussians is much
smaller than their separation $\alpha$, a fact which cannot be represented correctly in the figure.}
\label{fig3}
\end{center}
\end{figure}

To correctly account for an  experiment which embodies all fundamental aspects of the one under consideration,
one can choose a  wave function which exhibits entanglement of the $y_1$ and $y_2$ coordinates (we neglect the
$x$-coordinates which play no role for the argument), in which  the superposed factorized wave functions (any
finite number of them) should describe particles which, contrary to the tacit assumption made by Popper, are not
perfectly correlated in position but only rather accurately. In order to keep at minimum the level of
mathematical sophistication, we will consider the following initial wave function (see Fig.~\ref{fig3}):
\begin{equation}
\label{eq1}
\psi(y_1,y_2,t=0)=\frac{1}{\sqrt{3}}(\psi_{\alpha,\sigma}(y_1)\psi_{-\alpha,\sigma}(y_2)+\psi_{0,\sigma}
(y_1)\psi_{0,\sigma} (y_2) +\psi_{-\alpha,\sigma}(y_1)\psi_{\alpha,\sigma}(y_2))
\end{equation}
where $\psi_{\beta,\sigma}(y)=(2\pi \sigma^2)^{-1/4} {\textrm{exp}}(-\frac{(y-\beta)^{2}}{4\sigma^2})$ is a
Gaussian state with mean value  $\beta$ and standard deviation $\sigma$. We will also assume that $\alpha$ is
much larger than $\sigma$  and that, in turn, the initial width $\Delta$ of both slits is larger than
$2(\alpha+\sigma)$. Eq.~(\ref{eq1}) describes a pair of particles whose corresponding $y_i$-positions ($i=1,2$)
are  almost perfectly correlated (e.g., when the first particle is found to be within the interval
$[\alpha-2\sigma, \alpha+2\sigma]$ the other one will be found, almost with certainty, within
$[-\alpha-2\sigma,-\alpha+2\sigma]$) and whose one-particle wave functions $\psi_{-\alpha,\sigma},
\psi_{\alpha,\sigma}$ and $\psi_{0,\sigma}$ are almost orthogonal, since their centers are much more far away
from each other than their width, i.e., $\sigma \ll \alpha$. Of course, the parameter $\sigma$, which quantifies
the degree of localization, cannot be vanishingly small in order to satisfy the (above) request (ii), but, at
the same time, it is assumed to be sufficiently small to guarantee the desired quite accurate correlations in
position of the two particles.
At this point, if one narrows  the slit at R changing  its width from $\Delta$ to $\delta$, a quantity of the
order, let us say, of  $6\sigma$, essentially only the product state $\psi_{0,\sigma}(y_1)\psi_{0,\sigma}(y_2)$
will contribute to the activation of  the detectors placed behind the slit, since the other parts of  the wave
function $\psi(y_1,y_2,t=0)$ will be almost completely absorbed by the walls of the slit, yielding a negligible
contribution to the probability of activating some counters. One then has to take into account that,  given the
wave function $\psi_{0,\sigma}(y_1)\psi_{0,\sigma}(y_2)$, the particle at R has a probability approximately
equal to $99.7\%$ of being found within the interval $[-3\sigma, 3\sigma]$ of the $y_2$-axis which corresponds
to the opening of the slit. So, narrowing the slit at R  from $\Delta$ to   $\delta$ and considering only those
events in which detectors on both sides are activated, amounts to assume that the wave function of both
particles  is practically equal to $\psi_{0,\sigma}(y_1)\psi_{0,\sigma}(y_2)$.

We come now to show once more that this test cannot disprove the Copenhagen interpretation of quantum mechanics.
In order to reach such a conclusion, we let the wave function $\psi_{0,\sigma}(y_1)\psi_{0,\sigma}(y_2)$  evolve
freely for the time interval $(0,t)$,  which is the one needed to the particles moving along the $x$-axis to
cover the distance from the slits to the detectors. The evolved wave function remains a product of two gaussian
states, having the same mean value $y_1=y_2=0$ but with a larger standard deviation $\bar{\sigma}>\sigma$ equal
to:
\begin{equation}
\label{eq2} \bar{\sigma}=\sigma\sqrt{1+ \frac{\hbar^{2} t^{2}}{4m^{2}{\sigma}^{4}}},
\end{equation}
where $m$ is the mass of the particles. The standard deviation $\bar{\sigma}$ of Eq.~(\ref{eq2}) is a decreasing
function of $\sigma$ as long as $\sigma\in (0,\sqrt{\frac{\hbar t}{2m}})$: therefore, for $\sigma$ belonging to
such an interval,   the more localized the initial gaussian state is, the more it will spread in the time
interval $(0,t)$, as expected. So, in what follows, we will set $\sigma \simeq \sqrt{\frac{\hbar t}{2m}}$ in
order to get the minimum of $\bar{\sigma}$ and we assume that the physical parameters $m$ and $t$ are such that
the evolved wave function can  activate (with significant probability) only a limited fraction of all the
detectors (actually only those in front of the interval $[-\delta/2,\delta/2]$) at both sides. Now, in order to
reach the core of Popper's proposal, let us consider (with him) a different situation in which the width of the
slit at R is reduced to a value $d$ remarkably smaller than $\delta$ ($d\ll\delta=6\sigma$). This procedure
requires to change the wave function of the particles passing through the slit making it equal to
$N\psi_{0,\sigma}(y_1)\psi_{0,\sigma}(y_2)\chi_{d/2}(y_2)$, where $\chi_{d/2}(y_{2})$ is the normalized
characteristic function of the interval $[-\frac{d}{2},+\frac{d}{2}]$ and $N$ is a normalization factor. As it
is evident, this  narrowing of the slit has reduced to $d$ the support $\delta$ of the wave function at R but it
has left unchanged and equal to $\delta$ the one  of the  wave function at L. If we let the whole system evolve
for a time $t$, the standard deviation of the gaussian state at L remains that of Eq.~(\ref{eq2}) (that is, the
particle  at L will not increase its scatter with respect to the previously  considered case of an opening
$\delta=6\sigma$) while the corresponding particle at R will exhibit a considerably increased scatter.

To analyze this point in an extremely simplified but essentially correct way,  we suppose  that
$N\psi_{0,\sigma}(y_2)\chi_{d/2}(y_2)$ be approximately equivalent to the (normalized) characteristic function
$\chi_{d/2}(y_2)$ (this assumption is certainly correct since  $d\ll6\sigma$).

If we denote as $\phi(y_2,t)$ the wave function which is the evolved, in the time interval  $(0,t)$, of the
characteristic function $\chi_{d/2}(y_2)$, we obtain the following position   probability density~\cite{ghatak}
for the R particle:
\begin{equation}
\label{eq4} \vert\phi(y_2,t)\vert^{2}= \frac{1}{2d}[ (C(u+v) -C(u-v))^{2}+(S(u+v)-S(u-v))^{2}]
\end{equation}
where $u=y_{2}\sqrt{\frac{m}{\pi\hbar t}}$, $v=\frac{d}{2}\sqrt{\frac{m}{\pi\hbar t}}$ and
$C(\theta)=\int_{0}^{\theta}dz\cos({\frac{\pi}{2}z^{2}})$ and
$S(\theta)=\int_{0}^{\theta}dz\sin({\frac{\pi}{2}z^{2}})$ are the Fresnel integrals. The probability density  of
Eq.~(\ref{eq4}) exhibits the following simpler expression
\begin{equation}
\label{eq5} \vert\phi(y_2,t)\vert^{2} = \frac{2\hbar t}{md\pi} \frac{1}{y_{2}^{2}} \sin^{2}\Big(\frac{md}{2\hbar
t}\;y_{2}\Big)
\end{equation}
in the limiting case $v\ll 1$ (which amounts to $d\ll 2\sqrt{\frac{\pi \hbar t}{m}}$ or, equivalently, to $d\ll
5\sigma$). The function of Eq.~(\ref{eq5}) is the very well-known Fraunhofer diffraction pattern which describes
the intensity of a monochromatic light beam diffracted by a single slit~\cite{optics}. The relevant width
$\Delta y_{2}$ of such a curve is given by the distance between the two first minima (surrounding the absolute
maximum) and it  equals:
\begin{equation}
\label{eq6} \Delta y_{2}=\frac{4 \pi \hbar t}{md}.
\end{equation}
Now, if we choose $d=\frac{\sigma}{n}\simeq\frac{1}{n}\sqrt{\frac{\hbar t}{2m}}$, with $n$ integer arbitrarily
large, we see that the width of the Fraunhofer probability distribution $\Delta y_{2}$ increases arbitrarily
and, in particular, it becomes much greater than $6\bar{\sigma}$, the interval which defines the range of
counters which are activated with high probability ($>97\%$) whenever the slit at R is large enough to let the
whole wave function $\psi_{0,\sigma}$ to pass through. More precisely, one can prove that $\frac{\Delta
y_{2}}{6\bar{\sigma}} \approx\mathcal{O}(n)$.

Thus, as expected by Popper and as implied by the quantum rules, by narrowing the slit at R (i.e., by choosing
any $n > 2$) we may considerably increase the $y$-scatter of the corresponding particle of a factor $n$ but,
contrary to Popper's expectations and in full agreement with the quantum rules about wave packet reduction, we
leave unaffected  the wave function at L, which will then not exhibit an increased scatter.

Finally, if we look at Eq.~(1) of Ref.~\cite{collett} we easily understand why the formula given by the authors
conflicts (instead of being a consequence of) with the standard interpretation of quantum mechanics. In fact,
the considered formula pretends to relate the standard deviation $\Delta_{L}$ of the particle at L with the
width $s_{R}$ of the slit at R in the physical situation where the  slit at R has been narrowed while the slit
at L is left wide open, as follows:
\begin{equation}
\label{eq0.5} \Delta^{2}_{L}=\Big(\frac{d+r}{d}s_{R}\Big)^{2} +\Big(\frac{r\lambda}{4\pi s_{R}}\Big)^{2}
\end{equation}
where, according to the notation of Ref.~\cite{collett}, $\lambda$ is the particle wavelength, $d$ is the
distance from the source to the slits and $r$ is the distance from the slits to the detectors. This incorrect
formula implies that, whenever $s_{R}<[r/(d+r)]^{1/2}$, the second term dominates over the first and,
consequently, yields a larger scatter $\Delta_{L}$ for a smaller $s_{R}$.

 But, as it emerges clearly from our analysis, these two quantities are
completely uncorrelated and, consequently, all calculations performed in Ref.~\cite{collett} are irrelevant for
Popper's proposal, curiously enough just because the authors explicitly agree with Popper's (wrong) belief of
what quantum mechanics would predict for such an experiment.


\section{Conclusions}

In this paper we have illustrated the reasons why Popper's proposal~\cite{popper} cannot be considered neither a
crucial test for the Copenhagen interpretation of quantum mechanics nor for the principle of causality of
special relativity. More precisely, exhaustive arguments have been exhibited which explicitly show why one
cannot obtain a larger scatter of a particle at L by local actions (such as by narrowing a slit) on a particle
at R, even when the two particles are described by a (quite strictly spatially correlated) wave function.



\end{document}